\documentclass{aa}

\usepackage{graphicx}
\usepackage{txfonts}
\DeclareRobustCommand{\VAN}[3]{#2}
\let\VANthebibliography\thebibliography
\def\thebibliography{\DeclareRobustCommand{\VAN}[3]{##3}\VANthebibliography}

\usepackage{color}

\usepackage{textcomp}
\usepackage{amsmath}	
\usepackage{amssymb}	
\begin{document}

   \title{Measuring cavity powers of active galactic nuclei in clusters using a hybrid X-ray--radio method}
   \subtitle{A new window on feedback opened by subarcsecond LOFAR-VLBI observations}
   \titlerunning{A new hybrid X-ray--radio method for measuring cavity powers of active galactic nuclei}

   \author{R. Timmerman\inst{1}\fnmsep\thanks{E-mail: rtimmerman@strw.leidenuniv.nl (RT)}
        \and
        R. J. van Weeren\inst{1}
        \and
        A. Botteon\inst{1}
        \and
        H. J. A. Röttgering\inst{1}
        \and
        B. R. McNamara\inst{2}
        \and
        F. Sweijen\inst{1}
        \and
        L. B{\^\i}rzan\inst{3}
        \and
        L. K. Morabito\inst{4,5}
    }

   \institute{Leiden Observatory, Leiden University, P.O. Box 9513, 2300 RA Leiden, The Netherlands
              \and
              Department of Physics and Astronomy, University of Waterloo, Waterloo, ON, Canada
              \and
              Hamburg Observatory, Hamburg University, Gojenbergsweg 112, 21029 Hamburg, Germany
              \and
              Centre for Extragalactic Astronomy, Department of Physics, Durham University, Durham DH1 3LE, UK
              \and
              Institute for Computational Cosmology, Department of Physics, University of Durham, South Road, Durham DH1 3LE, UK
    }

   \date{Received XXX; accepted YYY}

 
  \abstract{Measurements of the quantity of radio-mode feedback injected by an active galactic nucleus into the cluster environment have mostly relied on X-ray observations, which reveal cavities in the intracluster medium excavated by the radio lobes. However, the sensitivity required to accurately constrain the dimensions of these cavities has proven to be a major limiting factor and is the main bottleneck on high-redshift measurements. We describe a hybrid method based on a combination of X-ray and radio observations, which aims to enhance our ability to study radio-mode feedback. In this paper, we present one of the first samples of galaxy clusters observed with the International LOFAR Telescope (ILT) at 144 MHz and use this sample to test the hybrid method at lower frequencies than before. By comparing our measurements with results found in literature based on the traditional method using only X-ray observations, we find that the hybrid method provides consistent results to the traditional method. In addition, we find that the correlation between the traditional method and the hybrid method improves as the X-ray cavities are more clearly defined. This suggests that using radio lobes as proxies for cavities may help to circumvent systematic uncertainties in the cavity volume measurements. Encouraged by the high volume of unique ILT observations successfully processed, this hybrid method enables radio-mode feedback to be studied at high redshifts for the first time even for large samples of clusters.}


   \maketitle
%

\section{Introduction}

The feedback process between active galactic nuclei (AGNs) and the dilute intracluster medium (ICM) that permeates a galaxy cluster is understood to be of critical importance in the formation and evolution of clusters of galaxies \citep[e.g.,][]{mcnamara07, fabian12, gitti12}. The hot ICM (\(T=10^7\,-\,10^8\)\,K) emits strong X-ray emission through thermal bremsstrahlung, which cools down this medium and causes it to sink down the gravitational well in the form of a cooling flow \citep{fabian94}. As this cooling flow accretes onto the brightest cluster galaxy (BCG) in the center of a cluster, it is expected to trigger a high rate of star formation. However, the observed star-formation rate is generally found to be lower than predicted based on the strength of the cooling flow \citep[e.g.,][]{fabian82, mcnamara89, kaastra01, peterson03, peterson06}. In addition, the amount of energy radiated away by the ICM suggests that this medium should cool down on a timescale much shorter than the lifetime of the cluster.

As the cooling flow accretes onto the BCG, it feeds the AGN located in the center of the galaxy. The resulting activity of the supermassive black hole produces large amounts of energy in the form of radiation and strong jetted outflows. In the scenario where the feedback is predominantly radiative, the AGN is said to be in a "quasar mode", whereas if the jetted outflows are dominant, the AGN is said to be in "radio mode" \citep{croton06}. As this energy is injected into the ICM, it completes the feedback cycle which prevents a runaway cooling event. The jetted outflows expand into large lobes against the internal pressure of the ICM \citep{bridle84}. This process can be observed in the radio regime of the electromagnetic spectrum, which shows the magnetized plasma emitted by the supermassive black hole, or using X-ray radiation, where these lobes produced by the AGN appear as cavities in the ICM \citep[e.g.,][]{bohringer93, carilli94}.

Efforts have been made to employ radio observations to study the radio-mode feedback of AGNs, but have not been able to demonstrate a reliable ability to measure the quantity of feedback \citep[e.g.,][]{birzan08, cavagnolo10, osullivan11, kokotanekov17}. Some of the best results have been obtained using bolometric radio luminosity measurements of the radio lobes, excluding the central core \citep{birzan04, birzan08}. However, this method has only obtained a weak correlation with X-ray measurements of the power required to excavate a cavity for two reasons. First of all, bolometric radio luminosity measurements require deep observations along a wide frequency range, although at high frequencies the sensitivity to the radio lobes suffers from the increasingly steep spectrum of the aging radio plasma. This makes bolometric radio measurements of radio lobes difficult to obtain. In addition, even with perfect bolometric radio luminosity measurements, the correlation with the X-ray cavity power is weakened by the dependence of the radio emissivity and X-ray cavity power on additional parameters. Most notably, the synchrotron emissivity depends on the square of the local magnetic field strength, and the particle composition of the jets determines the ratio between the total momentum of the jet and the momentum of the synchrotron-emitting electrons. Simulations show that the radio luminosity of lobes can vary with more than an order of magnitude for a given jet power \citep{hardcastle13, hardcastle14}.

The cavity power estimates derived from X-ray observations have generally been considered to offer the most reliable estimate of the amount of feedback \citep[e.g.,][]{mcnamara12, birzan04, birzan14}, but this method is also not without weaknesses. Clearly detecting the cavities requires very sensitive observations, which are infeasible at relatively high redshifts \citep[\(z>0.6\),][]{hlavacek15}. In addition, unless a cluster is very relaxed, the ICM will feature more structure than just the cavities, which can affect the ability to reliably constrain the shape and size of the cavities.

Recent developments in data processing \citep{morabito21} have enabled the calibration and imaging of observations taken with the complete International LOw Frequency ARray \citep[LOFAR,][]{haarlem13} Telescope (ILT). Long-baseline observations can now be reliably processed, even for complex \citep[e.g.,][]{timmerman21} or faint sources. The international stations of LOFAR are able to overcome the main obstacle of low-frequency radio observations: angular resolution. While the Dutch section of LOFAR offers an angular resolution of \(\theta\approx6~\mathrm{arcseconds}\) at 144~MHz, the inclusion of the international stations enables an angular resolution of \(\theta\approx0.3~\mathrm{arcseconds}\) to be reached. By observing at low frequencies with very long baselines, LOFAR is able to provide both the sensitivity and the angular resolution required to resolve the steep-spectrum emission from the radio lobes of AGNs.

We have observed a sample of cool-core galaxy clusters with AGNs up to a redshift of \(z=0.6\) using the ILT. Leveraging LOFAR's upgraded capabilities, we present a hybrid method of measuring the cavity power based on a combination of X-ray and radio data, which is likely to make these measurements feasible at higher redshifts than before. We test this method on our observed low-redshift sample to confirm that this method provides consistent results to those found in literature based on the traditional X-ray-based method.

In this paper, we adopt a \(\Lambda\)CDM cosmology with a Hubble parameter of \(H_0\ =\ 70~\mathrm{km}\ \mathrm{s}^{-1}\ \mathrm{Mpc}^{-1}\), a matter density parameter of \(\Omega_m\ =\ 0.3\), and a dark energy density parameter of \(\Omega_\Lambda\ =\ 0.7\). We define our spectral indices \(\alpha\) according to \(S \propto \nu^\alpha\), where \(S\) is flux density and \(\nu\) is frequency. All uncertainties denote the 68.3\%=1\(\sigma\) confidence interval.

\begin{table*}
    \caption{Summary of the sample of galaxy clusters used in this paper. The horizontal line separates the clusters that are studied at high resolutions ({\(\sim\)}0.3~arcseconds, top half) from the clusters that are studied at low resolutions ({\(\sim\)}6~arcseconds, bottom half).}
    \centering\small
    \begin{tabular}{lllllll}\hline\hline
        Cluster name & R.A. (J2000) & Dec. (J2000) & Redshift \(z\) & Mass M\(_{200}\)\\
                     &              &              &                & (\(10^{14}\) M\(_\odot\))\\\hline
        Perseus (NGC 1275) & 03h19m48.16s & 41d30m42.11s & 0.018 & \(6.65^{+0.43}_{-0.46}\) (1) \\
        Abell 1795 & 13h49m00.50s & 26d35m07.00s & 0.063 & \(15.39^{+3.17}_{-2.92}h_{50}^{-1}\) (2)\\
        Abell 2029 & 15h10m58.70s & 05d45m42.00s & 0.077 & \(11.2 \pm 0.4\) (3)\\
        ZwCl 2701 & 09h52m49.10s & 51d53m05.00s & 0.214 & \(5.2 \pm 0.5\) (3)\\
        4C+55.16 & 08h34m54.90s & 55d34m21.07s & 0.241 & \(0.923 \pm 0.182\) (4)\\
        RX J1532.9+3021 & 15h32m53.78s & 30d20m59.41s & 0.363 & \(11.1 \pm 1.1\) (3)\\
        MACS J1720.2+3536 & 17h20m16.80s & 35d36m27.00s & 0.391 & \(8.1 \pm 1.3\) (3)\\
        MACS J1423.8+2404 & 14h23m47.70s & 24d04m40.00s & 0.545 & \(7.2 \pm 0.9\) (3)\\\hline
        Perseus (NGC 1275) & 03h19m48.16s & 41d30m42.11s & 0.018 & \(6.65^{+0.43}_{-0.46}\) (1)  \\
        Abell 2199 & 16h28m37.00s & 39d31m28.00s &  0.030 & \(6.0 ^{+1.5}_{-1.8}\) (5,6) \\
        2A 0335+096 & 03h38m35.30s & 09d57m54.00s & 0.035 & \(1.4 ^{115.5}_{-1.0}\) (5,7)\\
        MKW 3S & 15h21m50.70s & 07d42m18.00s & 0.045 & \(5.1\) (8) \\
        Abell 1668 & 13h03m44.90s & 19d16m37.20s & 0.064 & \(4.31 \pm 1.63\) (9) \\
        Abell 2029 & 15h10m58.70s & 05d45m42.00s & 0.077 & \(11.2 \pm 0.4\) (3)\\
        3C 388 & 18h44m02.42s & 45d33m29.81s & 0.092 & ... \\
        MS 0735.6+7421 & 07h41m40.30s & 74d14m58.00s & 0.216 & \(7.12 \pm 1.96\) (10) \\\hline
    \end{tabular}
    \label{tab:sample}
    \tablebib{(1) \citet{simionescu11}; (2) \citet{reiprich02}; (3) \citet{mantz16}; (4) \citet{takey11}; (5) \citet{comerford07}; (6) \citet{lokas06}; (7) \citet{voigt06}; (8) \citet{pinzke11};  (9) \citet{kopylova17}; (10) \citet{gitti07}}
\end{table*}

\section{Methodology}

\subsection{Current cavity power estimation}

The most favored method to measure the amount of radio-mode feedback injected into the ICM by the AGN has been to perform a dynamical analysis on the cavities in the ICM. Following the method described by \citet{birzan04}, the amount of energy required to inflate a radio lobe (\(E_\text{cav}\)) is the sum of the internal energy of the lobe (\(E_\text{internal}\)) and the work required to excavate the region against the external ICM pressure (\(W\)). This gives

\begin{align}
    E_\text{cav} & = E_\text{internal} + W\notag\\
    & = \frac{1}{\gamma-1}pV + pV\\
    & = 4pV\notag
\end{align}

\noindent where \(\gamma\) is the adiabatic index of the radio lobe, \(p\) is the pressure of the surrounding ICM and \(V\) is the volume of the cavity. For a relativistic gas we know that \(\gamma=4/3\), leading to the final result.

To obtain the average power output of the AGN, this energy must be divided by the age of the cavity, which can be estimated in multiple ways. Firstly, because the radio lobe has a lower density than the surrounding ICM, a buoyant force acts upon this radio lobe, causing it to rise away from the center of the cluster. Assuming this dominates the kinematics of the lobe, the time required to reach the observed position of the cavity (the buoyancy timescale) can be estimated as

\begin{equation}
    t_\text{buoy} = R\sqrt{\frac{SC}{2gV}}
\end{equation}

\noindent where \(R\) is the projected distance to the center of the cluster, \(S\) is the cross-sectional area of the cavity, \(C\) is the drag coefficient \citep[\(C=0.75\),][]{churazov01}, and \(g\) is the gravitational acceleration which we estimate following \citet{birzan04} using the approximation of an isothermal sphere such that \(g\approx2\sigma^2/R\) \citep{binney01}, where \(\sigma\) is the stellar velocity dispersion.

Alternatively, because the lobe originates from a relativistic jet and therefore has a high amount of initial momentum, it can also be assumed that the cavity travels at the local speed of sound through the ICM. In this case, the age of the cavity can be estimated as

\begin{equation}
    t_\text{sound} = R\sqrt{\frac{\mu m_H}{\gamma kT}}
\end{equation}

\noindent where \(\mu\) is the mean molecular weight, \(m_H\) is the mass of a hydrogen atom, \(\gamma\) is again the adiabatic constant which we assume to be equal to 5/3 for the ICM, and \(kT\) is the thermal energy of the ICM. Generally, these two estimates agree within a factor of two for cavities close to the center of the cluster \citep[e.g.,][]{birzan04}, and diverge as the cavities are more distant to the AGN. As the buoyancy timescale is expected to be more accurate for older cavities and agrees reasonably well with the sound speed timescale for newer cavities, we adopt the buoyancy timescale for the remainder of this paper.

Deriving most of the quantities required for these calculations has generally been performed based on X-ray observations. Estimating the cavity energy requires the ICM pressure and the volume of the cavities. The ICM pressure can be derived, for instance, by determining the temperature and density of the ICM as a function of radius, which immediately provides an estimate of the pressure. To obtain the cavity volume, an estimate of the smooth brightness profile of the X-ray emission is subtracted from the image, causing the cavities to appear as negatively bright regions. From this, the dimensions and position of the cavity are generally derived assuming an ellipsoidal shape. Accurately measuring the cavity properties requires high-resolution X-ray observations, and therefore is almost exclusively performed using Chandra observations. These few quantities provide almost all the information required to estimate the cavity power, with only the stellar velocity dispersion remaining unknown in the case that the buoyancy timescale is assumed. This final parameter is normally derived through optical spectroscopy.

\subsection{The hybrid X-ray--radio method}

Although the high angular resolution provided by Chandra in principle enables small spatial scales to be resolved even at high redshifts, X-ray observations are plagued by a relatively low count statistics at high redshifts, which forms a bottleneck for the purely X-ray-based approach to constrain the size and shape of cavities. In addition, the ICM is in general not a smooth distribution. Additional structure in the ICM due to, e.g., a recent merging event can make it difficult to reliably identify the cavities.

The radio lobes have previously been treated as proxies for the cavities in select clusters \citep[e.g.,][]{dunn04,allen06,cavagnolo10,gitti10,lanz10,ehlert11,vagshette17,seth22}, though generally this is avoided at high frequencies due to the lack of sensitivity to low-energy cosmic ray electrons, resulting in high systematic uncertainties. The first detailed study of this method was performed by \citet{birzan08} by comparing the cavity enthalpy estimated based on 327 MHz and 1400 MHz Very Large Array (VLA) observations with measurements obtained from Chandra X-ray images, where they found the 327 MHz observations to consistently overestimate the cavity enthalpy. Following this test, we employ radio observations taken with the ILT for the first time to probe the position, shape and size of the X-ray cavities by treating the radio lobes as proxies for the cavities. The enthalpy of the cavities is then derived using the volume measurements derived from radio observations and a pressure measurement derived from X-ray observations. Thanks to LOFAR's international stations, we now have a unique combination of angular resolution and sensitivity, offering an unprecedented view of the radio lobes.

This method has the advantage that in many instances the radio lobes will be detected much more clearly than the X-ray cavities, enabling cavity powers to be derived at higher redshifts than before. In addition, this method only requires observations at a single frequency, which therefore will also offer results more robustly compared to the multi-frequency observations which were previously used to calculate the bolometric radio luminosity of the lobes. It is not uncommon for radio lobes to only be detected at low frequencies due to their steep spectra. Also, this method conveniently avoids requiring accurate flux scale calibration, which is known to be one of the main weaknesses of the current ILT calibration pipeline. Finally, because this method only depends on X-ray observations for ICM pressure measurements, which does not require high angular resolutions, it becomes more feasible for observatories like XMM-Newton to assist with the cavity power measurements.

We note that one of the main sources of uncertainty from the purely X-ray-based approach persists in this method: the unknown projection effects. In addition, the sensitivity to the steep-spectrum radio plasma will decrease as this plasma ages, causing very old radio lobes to remain undetected in a standard-depth LOFAR pointing. It should be carefully checked if there is likely to be radio emission below the detection limit. If so, the radio-derived volume estimates may be unreliable. Finally, we note that even though it may be tempting to derive the equipartition pressure of the lobes, these estimates in general do not agree with the ICM pressure derived from X-ray observations \citep{croston14}.

\section{Sample}\label{sec:sample}

We have compiled a sample of 8 targets based on the samples of \citet{rafferty06} and \citet{birzan08} for new high-resolution observations and a sample of 8 targets for low-resolution comparison based on the samples of \citet{rafferty06} and \citet{birzan20}. Two targets of the low-resolution sample are in common with the high-resolution sample. This collective sample of 14 targets was based on the detection of X-ray cavities with an associated radio source, and spans a redshift range from \(z=0\) up to \(z=0.6\) (see Table \ref{tab:sample}).

For our high-resolution sample, we have processed and imaged observations taken with the ILT as described in Sect. \ref{sec:lofardatareduction}. For our low-resolution sample, we mainly depend on images published as part of the LOFAR Two-Metre Sky Survey \citep[LoTSS,][]{shimwell16,shimwell19, shimwell22}. For 2A\,0335+096 and MS\,0735.6+7421, LoTSS images are not available, so we instead use the images published in \citet{ignesti21} and \citet{biava21}, respectively. Finally, for Perseus we have produced a low-resolution image using only the Dutch LOFAR stations after subtracting the dominant central component as observed in the high-resolution map, which previously interfered with attempts to image the cluster with an angular resolution of \({\sim}6\) arcseconds.

In addition to the LOFAR observations, we have processed archival VLA observations of our high-resolution sample to produce spectral index maps and aid with the identification of the radio structures. The details of these observations and the data reduction are described in Appendix \ref{appendix:vla}.

\section{Observations and data reduction}

\begin{table*}
    \caption{Summary of the LOFAR observations processed for the high-resolution images presented in this paper.}
    \centering\small
    \begin{tabular}{lllll}\hline\hline
        Cluster name & Project code & PI & Date & Duration\\\hline
        Perseus (NGC 1275) & LC6\_015 & Shimwell & 3 Nov 2016 & 8 hours \\
        Abell 1795 & LC7\_024 & Shimwell & 9 Feb 2017 & 8 hours\\
        Abell 2029 & LC14\_019 & Timmerman & 10 Sep 2020 & 4 hours\\
                   &           &           & 4 Oct 2020 & 4 hours\\
        ZwCl 2701 & LC9\_019 & B{\^\i}rzan & 20 Feb 2018 & 8 hours\\
        4C+55.16 & LC14\_019 & Timmerman & 9 Nov 2020 & 8 hours\\
        RX J1532.9+3021 & LC10\_010 & Bonafede & 14 Sep 2018 & 8 hours\\
        MACS J1720.2+3536 & LC10\_010 & Bonafede & 9 Jun 2018 & 8 hours\\
        MACS J1423.8+2404 & LC14\_019 & Timmerman & 8 Oct 2020 & 8 hours\\\hline
    \end{tabular}
    \label{tab:LOFAR}
\end{table*}

\subsection{LOFAR}\label{sec:lofardatareduction}

The sources in our sample have been observed with LOFAR's High Band Antennas at frequencies between 120~MHz and 168~Hz for a total of 8~hours per target. Each target observation was preceded and succeeded by a 10-minute long observation of a well-known calibrator source for gain and bandpass calibration purposes. The initial data reduction and calibration was performed using the \textsc{Prefactor} software package \citep{vanweeren16, williams16, gasperin19}. After completing the initial flagging of data \citep{offringa13, offringa15}, \textsc{Prefactor} derived the calibration solutions based on the calibrator source. These calibration solutions consist of the corrections for the polarization alignment and Faraday rotation, the bandpass, and the clock offsets. After applying all calibration solutions to the data, \textsc{Prefactor} performed another round of flagging and averaged the data to 8~seconds per integration and frequency channels with a bandwidth of 98~kHz. As a final step, a sky model of the field as provided by the TIFR Giant Metrewave Radio Telescope Sky Survey \citep[TGSS,][]{intema17} is used to calibrate the phases of the visibilities for the Dutch stations.

With the initial calibration of the Dutch stations completed, the LOFAR-VLBI pipeline developed by \citet{morabito21} is used to extend the calibration to the international LOFAR stations. First, the previously derived calibration solutions are transfered and applied to the target observation. Next, we select a bright and compact source in the field from the Long-Baseline Calibrator Survey \citep[LBCS,][]{jackson16, jackson21}, and use this to calibrate the international stations. In the case of Abell 2029, no calibrator source in the field was known, so one had to be manually found. After obtaining the calibration solutions for the international stations on the calibrator source, the solutions are transferred to the target source.

Due to the direction dependence of the calibration solutions, additional calibration has to be performed on the target source itself after transferring the previously derived solutions. To perform this final self-calibration \citep{vanweeren21}, the Default Preprocessing Pipeline \citep[\textsc{DPPP},][]{vandiepen18} was employed to derive and apply updated calibration solutions, and \textsc{WSClean} \citep{offringa14} was employed to produce an image of the source. The self-calibration consisted of total electron content (TEC) and phase corrections, and depending on the brightness of the target source also of amplitude corrections. The core stations of LOFAR were combined into a single virtual station to narrow down the field of view and reduce interference from unrelated sources near the target. The angular resolution of the final images is on the order of 0.3~arcseconds at a central frequency of 144~MHz, with small variations between the different observations.

\subsection{Chandra}

The superb sub-arcsecond resolution of the \textit{Chandra} satellite provides a good match to that achieved by LOFAR-VLBI observations. For this reason, we made use of archival \textit{Chandra} Advanced CCD Imaging Spectrometer (ACIS) data to study the X-ray cavities associated with the AGN in our sample. Data were retrieved from the \textit{Chandra} data archive\footnote{\url{https://cda.harvard.edu/chaser}} and processed with \textsc{ciao} v4.12 and CALDB v4.9.0 starting from the \texttt{level=1} event file. Observing periods affected by background flares were removed by inspecting light curves extracted in the 0.5$-$7.0 keV band using the \texttt{deflare} task. When multiple ObsIDs were available, event files were combined with the \texttt{merge\_obs} script. All images used for the analysis are exposure-corrected and were obtained in the 0.5$-$2.0 keV band. A summary of the ObsIDs used in this work together with the total net exposure time per cluster is reported in Table~\ref{tab:chandradata}.

\begin{table*}
    \caption{Summary of the \textit{Chandra} observations processed for the X-ray residual maps presented in this paper.}
    \centering
    \begin{tabular}{lll}\hline\hline
        Cluster name & Net exposure time & Observation ID\\
        & (ks) & \\\hline
        Perseus (NGC 1275) & 1450 & 1513, 3209, 4289, 4946...4953, 6139, 6145, 6146, 11713...11716, 12025, 12033,\\
        & & 12036, 12037 \\
        Abell 2199 & 156 & 497, 498, 10748, 10803...10805\\
        2A 0335+096 & 82 & 7939, 9792\\
        MKW 3S & 56 & 900\\
        Abell 1795 & 1161 & 493, 494, 3666, 5286...5289, 6160, 6162, 6163, 10900, 12026, 12028, 12029,\\
                   &      & 13106...13108, 13110, 13414, 14268..14270, 15485...15487, 16432...16434,\\
                   &      & 16436, 16465, 17228, 17397...17399, 17401, 17405, 17408, 17683, 17685,\\
                   &      & 17686, 18423...18427, 18429, 18433, 19868...19870, 19877...19879, 19968,\\
                   &      & 19969, 20642...20644, 20651...20653, 21830...21832, 21839...21841,\\
                   &      & 22829...22831, 22838...22840, 24602, 24609...24611\\
        Abell 1668 & 10 & 12877\\
        Abell 2029 & 126 & 891, 4977, 6101, 10434...10437\\
        3C 388 & 35 & 4756, 5295\\
        ZwCl 2701 & 127 & 3195, 7706, 12903\\
        MS 0735.6+7421 & 512 & 4197, 10468...10471, 10822, 10918, 10922\\
        4C+55.16 & 94 & 1645, 4940\\
        RX J1532.9+3021 & 107 & 1649, 1665, 14009\\
        MACS J1720.2+3536 & 60 & 3280, 6107, 7225, 7718\\
        MACS J1423.8+2404 & 133 & 1657, 4195\\\hline
    \end{tabular}
    \label{tab:chandradata}
\end{table*}

\section{Results}

\begin{figure*}
    \centering
    \includegraphics[width=\textwidth]{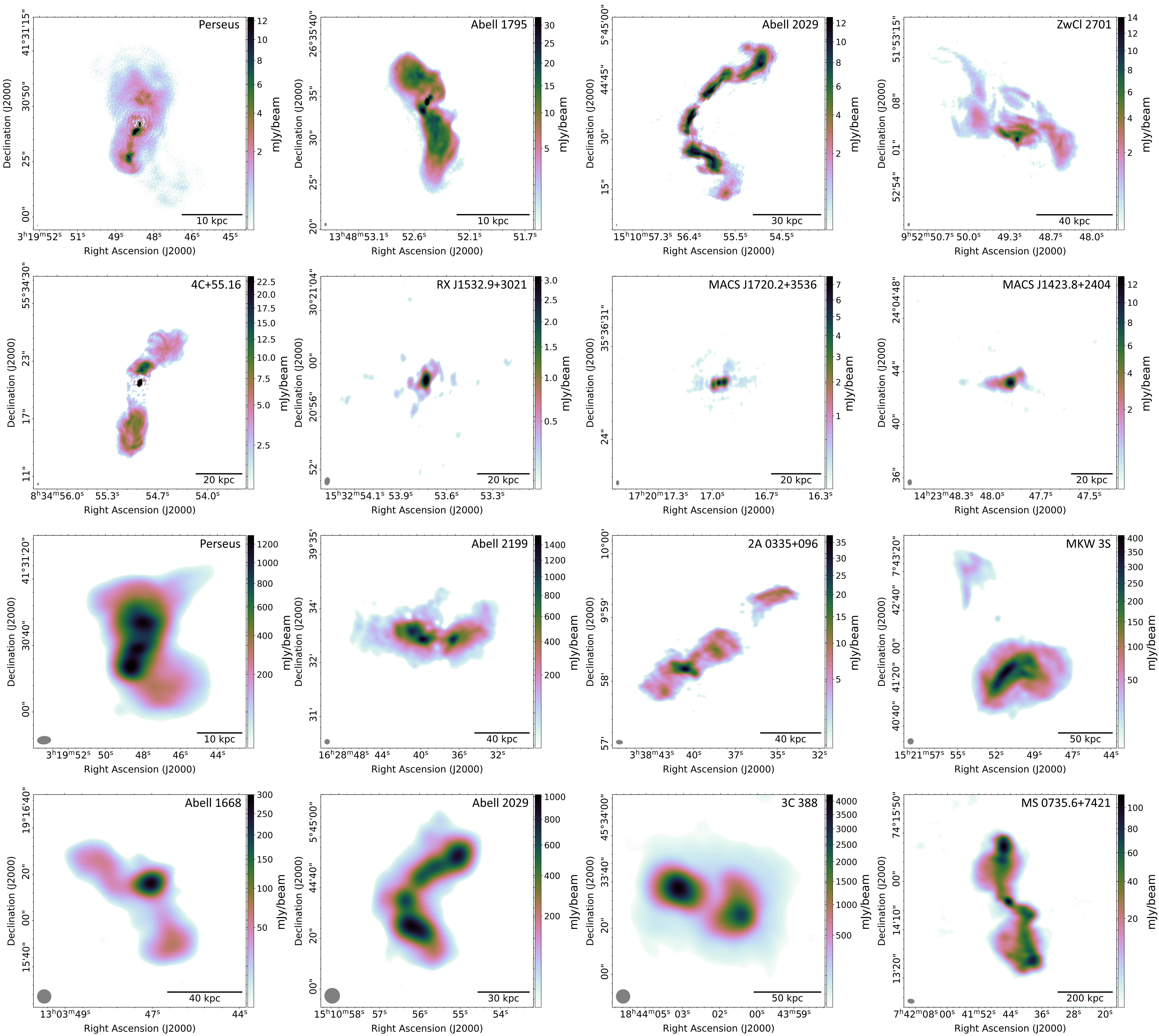}
    \caption{LOFAR images of all galaxy clusters our the sample. The top 8 panels show the high-resolution maps produced by including LOFAR's international stations and the bottom 8 panels show the low-resolution maps produced using only the Dutch part of the array. Note that Perseus and Abell\,2029 are presented in both sections. The central dominant compact component in the Perseus cluster was peeled from the data for the low-resolution map for calibration purposes. The color maps range from three times the rms noise level to the peak brightness, except in the cases of Perseus, Abell\,1795 and 4C+55.16, for which the peak brightness of the lobes was used due to the otherwise dominant central compact component. The scale bar in the bottom right corner of each panel measures the listed length at the redshift of the respective clusters. The beam is indicated in grey in the bottom left corner of each panel.}
    \label{fig:images}
\end{figure*}

\begin{figure*}
    \centering
    \includegraphics[width=\textwidth]{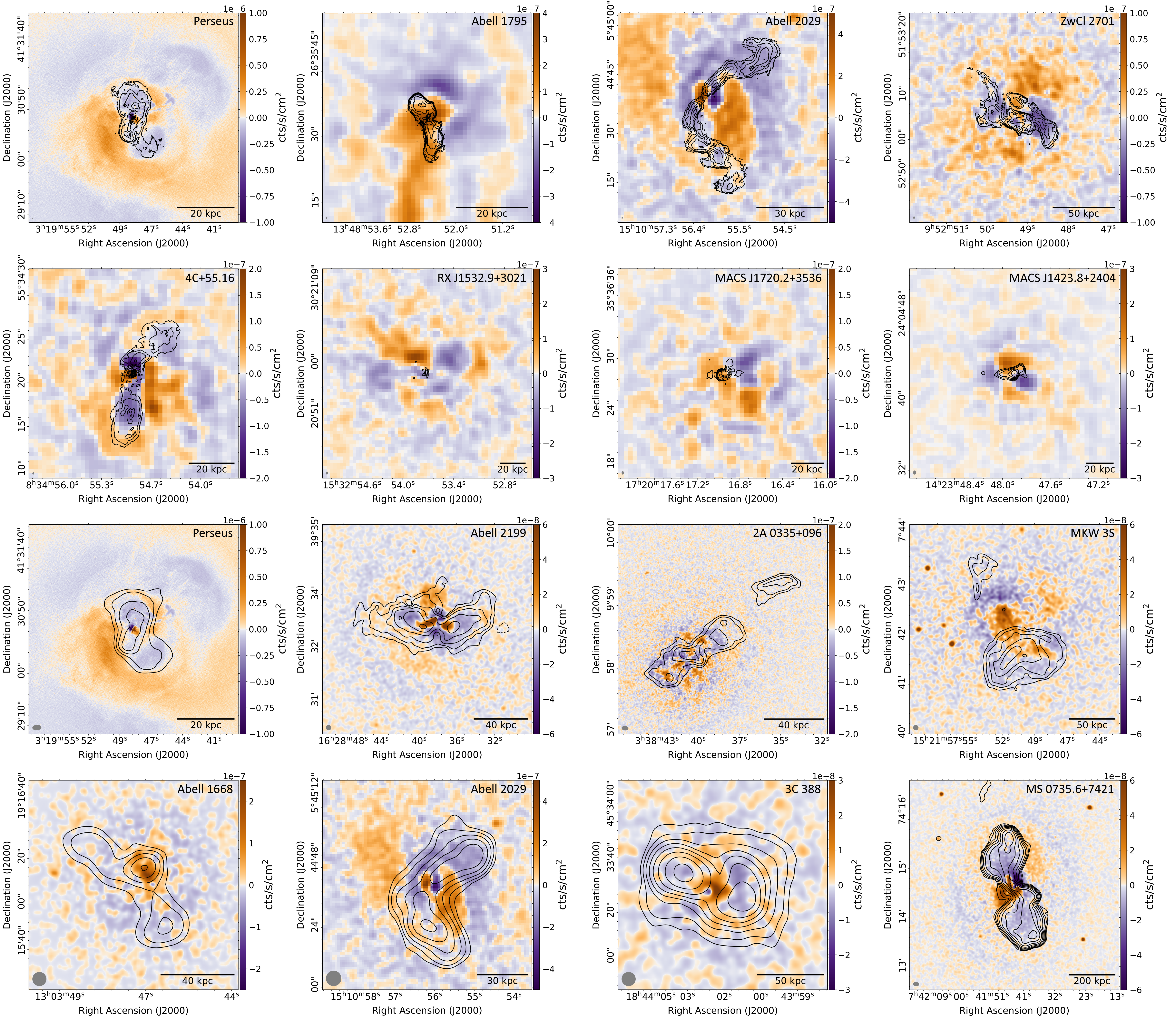}
    \caption{Residuals from the \textit{Chandra} X-ray images after subtracting a smooth model of the ICM surface brightness. The top 8~panels show the X-ray residual maps corresponding to the high-resolution sample and the bottom 8~panels show the X-ray residual maps corresponding to the low-resolution sample. Note that Perseus and Abell\,2029 are presented in both sections. The orange and purple colors indicate regions with surface brightness excess and deficiency, respectively. The black contours indicate the radio emission, and are drawn in increments of factors of 2, starting at 5~times the rms noise level. The scale bar in the bottom right corner of each panel measures the listed length at the redshift of the respective clusters. The beam is indicated in grey in the bottom left corner of each panel.}
    \label{fig:xrayoverlay}
\end{figure*}

\begin{figure}
    \centering
    \includegraphics[width=\columnwidth]{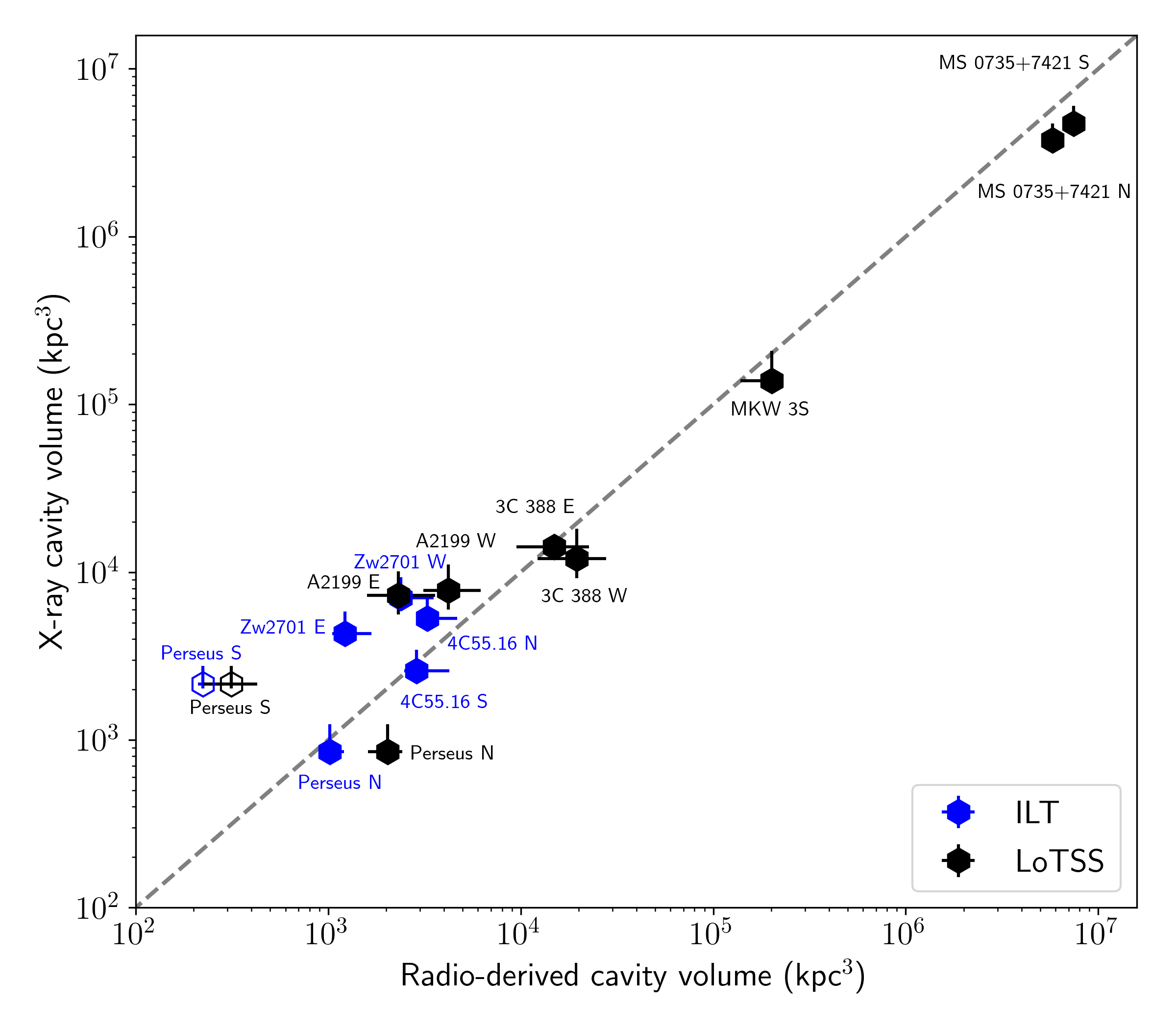}
    \caption{Radio-estimated cavity volumes versus X-ray-estimated cavity volumes. The blue data points indicate the measurements derived using high-resolution observations, while the black data points indicate the measurements derived using low-resolution observations. The dashed line indicates the line of equality. The open markers indicate the cavities for which the radio lobe does not visually match the cavity as observed in the X-ray.}
    \label{fig:sizesize}
\end{figure}

\begin{figure}
    \centering
    \includegraphics[width=0.7\columnwidth]{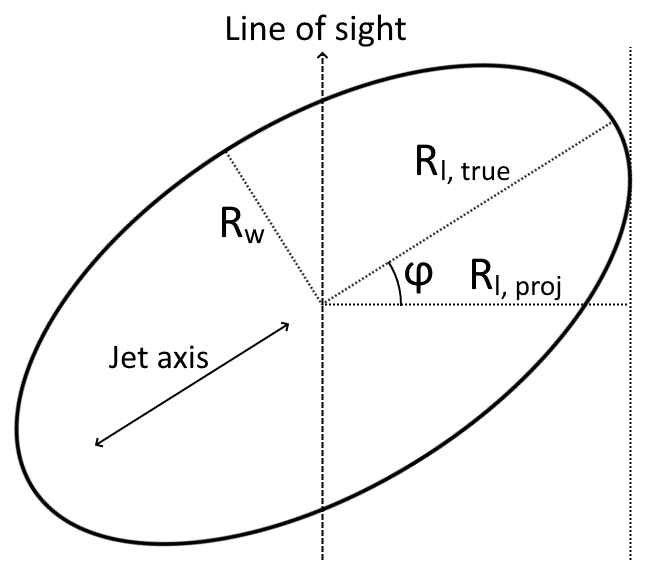}
    \caption{Schematic of the construction used to estimate the uncertainties on the cavity volume measurements.}
    \label{fig:schematic}
\end{figure}

\subsection{Imaging}

To study the radio lobes present in our sample, we have imaged all sources in our high-resolution (\({\sim}0.3\)~arcseconds) sample using WSClean with a Briggs robust parameter of -0.5 \citep{briggs95}, as shown in Fig.~\ref{fig:images}. The images of the low-resolution (\({\sim}6\)~arcseconds) sample have mainly been sourced externally, with the Perseus cluster forming the only exception, as previously discussed in Sect. \ref{sec:sample}.

After producing the X-ray images, a Multi-Gaussian Expansion \citep{cappellari02} was used to subtract a smooth model of the ICM brightness distribution from the images to obtain the residual map, where the cavities are most clearly visible. This technique is particularly efficient in revealing surface brightness depressions like cavities, as demonstrated by, e.g., \citet{rafferty13} and \citet{birzan20}. Depending on the total photon count, the residuals were smoothed with a Gaussian function to reduce noise. In the case where background structures obfuscate the cavities, an unsharp mask was applied to flatten out the image while maintaining the cavity structures. The LOFAR images were overlaid on the X-ray residuals to correlate the radio lobes with the X-ray cavities, as shown in Fig.~\ref{fig:xrayoverlay}.

\subsubsection{High-resolution sample}

In this section we briefly describe the high-resolution radio images (see Fig. \ref{fig:images}), radio/X-ray overlays (see Fig. \ref{fig:xrayoverlay}) and spectral index maps (see Fig. \ref{fig:spixmaps}) of individual clusters.

\begin{itemize}
    \item \textbf{Perseus}: The high-resolution image of Perseus shows a dominant flat-spectrum central compact component surrounded by diffuse emission. The steeper-spectrum (\(\alpha=-0.9\)) jetted outflows from the AGN are visible towards the north and the south. The lingering steep-spectrum (\(\alpha=-1.8\)) emission of a previous outburst can still be seen towards the southwest of the center, similar to as previously reported based on VLA and MERLIN observations by \citet{pedlar90} and \citet{gendronmarsolais17, gendronmarsolais20}. From the X-ray overlay, it is clear that all detected radio emission is coincident with an X-ray depression \citep{fabian00}. The X-ray residuals, the radio morphology, and the spectral indices strongly suggest that the AGN experienced at least two distinct outbursts.
    \item \textbf{Abell\,1795}: There are two clear radio lobes visible towards the north and the south. As previously observed with the VLA by \citet{breugel84}, it appears that the radio jet is emitted along the northeast to southwest direction, and both jets then bend away towards the north and south, respectively. The spectral index gradient along the lobe clearly shows the direction of the outflow. There is no clear correlation between these radio lobes and an X-ray depression, although this is likely to be at least in part due to the presence of a long X-ray filament stretching from the center of the cluster about 40~arcseconds towards the south \citep{fabian01}, which can affect both the cavity structure and our ability to detect it.
    \item \textbf{Abell\,2029}: The radio structure of Abell\,2029 as detected with LOFAR shows an FR~I-type morphology \citep{fanaroff74} with two radio filaments towards the northwest and the south, similar to as observed by \citet{taylor94} using the VLA. Similar as to in Abell 1795, the spectral index gradually steepens towards the older regions of the outflows. The radio lobes do not clearly coincide with depressions in the X-ray brightness, although the spiral pattern detected in the X-ray residuals \citep{clarke04} may hinder a clear detection of cavities in a comparable manner as with Abell\,1795. We also note the presence of multiple filamentary structures throughout the radio outflows.
    \item \textbf{ZwCl\,2701}: The low-frequency radio structure of ZwCl\,2701 shows complex features previously unreported in literature. A mildly bright compact component with a peak brightness of 14~mJy is located in the central region, surrounded by diffuse emission. An apparently separated filamentary structure is located approximately 6~arcseconds north of the core. Two radio lobes previously reported by \citet{birzan20} can be found approximately 7~arcseconds east and west of the core, where they coincide with the cavities previously also observed in the X-ray by \citet{vagshette16}. An additional tail-like structure is seen just north of the eastern lobe, but due to the morphology of this structure and the lack of coinciding X-ray brightness depression, it is assumed that this structure is not a constituent of the radio lobe.
    \item \textbf{4C+55.16}: The structure of 4C+55.16 consists of a bright compact flat-spectrum component in the center with two steeper-spectrum radio lobes extending about 9~arcseconds in the northwestern and southern directions, similar to as observed by \citet{xu95}. The southern lobe coincides with a very clear cavity with an X-ray bright rim almost fully surrounding the cavity. The northern radio lobe also coincides with a cavity, but the extent of this cavity is less pronounced \citep{hlavacek11}.
    \item \textbf{RX\,J1532.9+3021}: Only a faint compact radio component can be detected in the LOFAR map of RX\,J1532.9+3021, similar as to previously reported by \citet{yu18}. Although the X-ray residuals show two clear brightness depressions towards the east and west of the cluster \citep{hlavacek13}, no coincident radio emission is detected.
    \item \textbf{MACS\,J1720.2+3536}: The LOFAR map reveals three faint compact components located within one arcsecond of one another, consistent with observations by \citet{yu18} using the VLA. It is unclear what the physical nature is of these components, but it is not assumed that the two outer components represent radio lobes. \citet{hlavacek12} report the detection of a clear cavity north of the core and a fainter cavity towards the south, but no radio emission is found to be coincident with those regions.
    \item \textbf{MACS\,J1423.8+2404}: This cluster is the most distant object in our sample, with a redshift of \(z=0.545\) \citep{limousin10}. The structure of MACS\,J1423.8+2404 shows a central faint compact component with two elongations towards the east and northwest. A hint of this eastern extension was also reported by \citet{yu18} based on 1.5~GHz VLA observations. These extensions may be related to the radio jets, as two X-ray brightness depressions are detected roughly in the same location, but these can not be relied on to provide a description of the radio lobes due to their low significance.
\end{itemize}

\subsubsection{Low-resolution sample}

In this section we briefly describe the low-resolution radio images (see Fig. \ref{fig:images}) and overlays (see Fig. \ref{fig:xrayoverlay}) of individual clusters. Low-resolution LOFAR maps of Abell\,2199, 2A\,0335+096, MKW\,3S, Abell\,1668 and MS0735.6+7421 were previously published by \citet{birzan20}. In addition, the low-resolution maps presented in Fig. \ref{fig:images} of 2A\,0335+096 and MS\,0735.6+7421 were obtained from \citet{ignesti21} and \citet{biava21}, respectively. Finally, a low-resolution LOFAR map of 3C\,388 was already previously published by \citet{brienza20}.

\begin{itemize}
    \item \textbf{Perseus}: To aid with the calibration of the data and obtain a clear view of the diffuse emission, the central dominant compact component was peeled from the data based on the high-resolution imaging. The northern and southwestern lobes are still clearly resolved. The southwestern lobe is more prominently detected in the low-resolution map and completely fills the X-ray cavity.
    \item \textbf{Abell\,2199}: The LoTSS map reveals two extended radio lobes located towards the east and west of the core of the cluster, similar to as previously observed at 408~MHz with the One-Mile Telescope (OMT) at Cambrige by \citet{parker67}. Similar to as observed with the OMT, no central compact component can be detected in the LOFAR map, although this component was present at higher frequencies in previously reported Westerbork Synthesis Radio Telescope maps \citep{jaffe74} and VLA maps \citep{burns1983}, suggesting that the core is strongly self-absorbed. Despite the chaotic structure in the X-ray residuals, it is clear that the radio lobes coincide with the observed cavities, as also previously reported by \citet{johnstone02}.
    \item \textbf{2A\,0335+096}: As reported by \citet{ignesti21}, the radio structure of this cluster at 144~MHz consists predominantly of diffuse emission. Although no compact central component is detected, the central region shows a bright extended component. Two diffuse radio lobes are located towards the northwest and southeast of the center. A diffuse component is located further towards the northwest, and is apparently separated from the rest of the cluster. It is unclear whether this structure is a radio lobe associated with a previous outburst of the AGN, or if it has a different physical nature. The X-ray residuals show a chaotic structure \citep{mazzotta03, sanders09}, which hampers a clear identification of the cavities.
    \item \textbf{MKW\,3S}: Similar to as reported by \citet{mazzotta02}, a single bright extended radio lobe towards the south of the core dominates the radio structure of MKW\,3S. An additional fainter diffuse component is located towards the north, but no compact emission associated with the AGN is detected. The X-ray residuals show a cavity coincident with the southern radio lobe, but although there is a surface brightness depression towards the north, this is not found to be directly coincident with the northern radio emission.
    \item \textbf{Abell\,1668}: Two radio lobes towards the northeast and south are clearly detected in the LoTSS map. These radio lobes are directly connected to a central compact component. \citet{pasini21} report two possible X-ray cavities towards the northwest, but these do not coincide with the observed radio emission. The lack of clear cavity detections may in part be due to the low sensitivity of the available X-ray observations.
    \item \textbf{Abell\,2029}: From the low-resolution radio map of Abell\,2029 we see the same radio lobes towards the northwest and south of the core. However, whereas both our high-resolution map and the previously reported VLA maps \citep{clarke04} show more bent lobe structures, these features are not resolved in the low-resolution map. Similar as with the high-resolution map, the spiral pattern detected in the X-ray residuals hinders a clear identification of cavities coincident with the radio lobes.
    \item \textbf{3C\,388}: Similar to as observed with the OMT by \citet{mackay69} and \citet{branson72}, the LoTSS map shows two bright radio lobes located towards the east and west of the core of the cluster. No central compact component can be clearly detected. Both of the radio lobes are coincident with X-ray depressions \citep{kraft06}, although the low sensitivity of the X-ray observations does not enable these cavities to be studied in detail.
    \item \textbf{MS\,0735.6+7421}: As reported by \citet{biava21}, the LOFAR map of MS\,0735.6+7421 shows two bright radio lobes extending towards the north and south of a central compact component. This structure is mainly similar to as observed with the VLA by \citet{cohen05}. The X-ray residuals show cavities directly coincident with the entire extent of the radio emission, and feature clear rims fully surrounding the cavities \citep{mcnamara05}.
\end{itemize}

\subsection{Analysis}

From our LOFAR images, we measure the size of the radio lobes assuming an ellipsoidal shape. The major and minor axes of the radio lobes are estimated by eye, preferably based on the CLEAN models, to optimally incorporate the identification of the different components present in the images. In the situation where a radio lobe does not feature clearly-defined edges (e.g., 3C\,388), the dimensions are estimated based on the steepest gradients surrounding the lobe. Due to the low surface brightness, the lobes could not be reliably identified for three sources in our sample: RX\,J1532.9+3021, MACS\,J1720.2+3536 and MACS\,J1423.8+2404. For this reason, these three sources have been excluded from further analysis. This will be further discussed in Sect.~\ref{sec:performance}.

One of the most fundamental differences between the hybrid X-ray--radio method and the purely X-ray-based method is that the cavity volume is now derived based on radio observations. To confirm that these measurements are reliable, we compare the X-ray-derived estimates for the cavity volume, as found in the literature, to our radio-derived estimates, as shown in Fig.~\ref{fig:sizesize}. As the dominant uncertainty of the volume measurements is generally due to projection effects, we quantify this uncertainty through a Monte Carlo approach where we randomly select an orientation for the cavity and calculate which true volume would correspond to the projected dimensions of the cavity. Assuming a projected semi-major length along the jet axis \(R_\text{l, proj}\) and a projected semi-minor axis perpendicular to the jet axis \(R_\text{w}\), the deprojected semi-major axis of the cavity can be calculated as

\begin{equation}
    R_\text{l, true}\;=\;\sqrt{\frac{R_\text{l, proj}^2-R_w^2\sin^2\varphi}{\cos^2\varphi}}
\end{equation}

\noindent where \(\varphi\) is the angle between the true jet axis and the projected plane. The semi-minor axis of the cavity is independent of the orientation. Given the deprojected semi-major axis, the cavity volume can be calculated as \(V=\frac{4}{3}\pi R_\text{l, true}R_\text{w}^2\). A schematic of this construction is shown in Fig.~\ref{fig:schematic}. For consistency, we recalculate the uncertainties on the cavity volumes of the literature X-ray estimates as well using the same method. In addition, we implement an uncertainty on the projected dimensions of the cavity. For literature values, we adopt an uncertainty on the order of the most precise digit if the uncertainty is not published (e.g., an uncertainty of 1~kpc on 13~kpc and 0.1~kpc in the case of 13.0~kpc). For our radio estimates, we assume an uncertainty of a quarter of the synthesized beam width. The final cavity volume estimates are then calculated as the median of the cavity volume probability distribution function, with the 68.3\% confidence interval serving as the projection-based uncertainty.

In some cases, a direct comparison between the X-ray cavity and the observed radio lobes is not possible. For instance, \citet{rafferty06} only report one cavity in the case of Abell\,1795 and Abell\,2029, leaving it unknown which radio lobe should match this cavity. Based on the significant difference in the reported distance to the center between the X-ray cavity and the radio lobe, it is safe to conclude that these do not correspond to the same structure. Similarly, although two cavities are reported in 2A\,0335+096, the perturbed structure at the core hampers a certain match between the X-ray cavities and the radio lobes. Finally, in the case of Abell\,1668, cavities are detected towards the north and north-west of the center, whereas the radio lobes are detected at larger radii towards the south and north-east. Due to these issues, the aforementioned systems are excluded from Fig.~\ref{fig:sizesize}.

Following this comparison, we proceed by calculating the cavity power corresponding to the observed radio lobes. For this calculation, we follow the same Monte Carlo procedure as before, where we assume a random orientation and use the resulting true cavity volume and distance to the center of the cluster to calculate the cavity volume. For the ICM pressure, we assume the same values used in literature for the X-ray cavity power estimates and do not vary this with radius as the pressure profiles are generally not published. 

\begin{figure}
    \centering
    \includegraphics[width=\columnwidth]{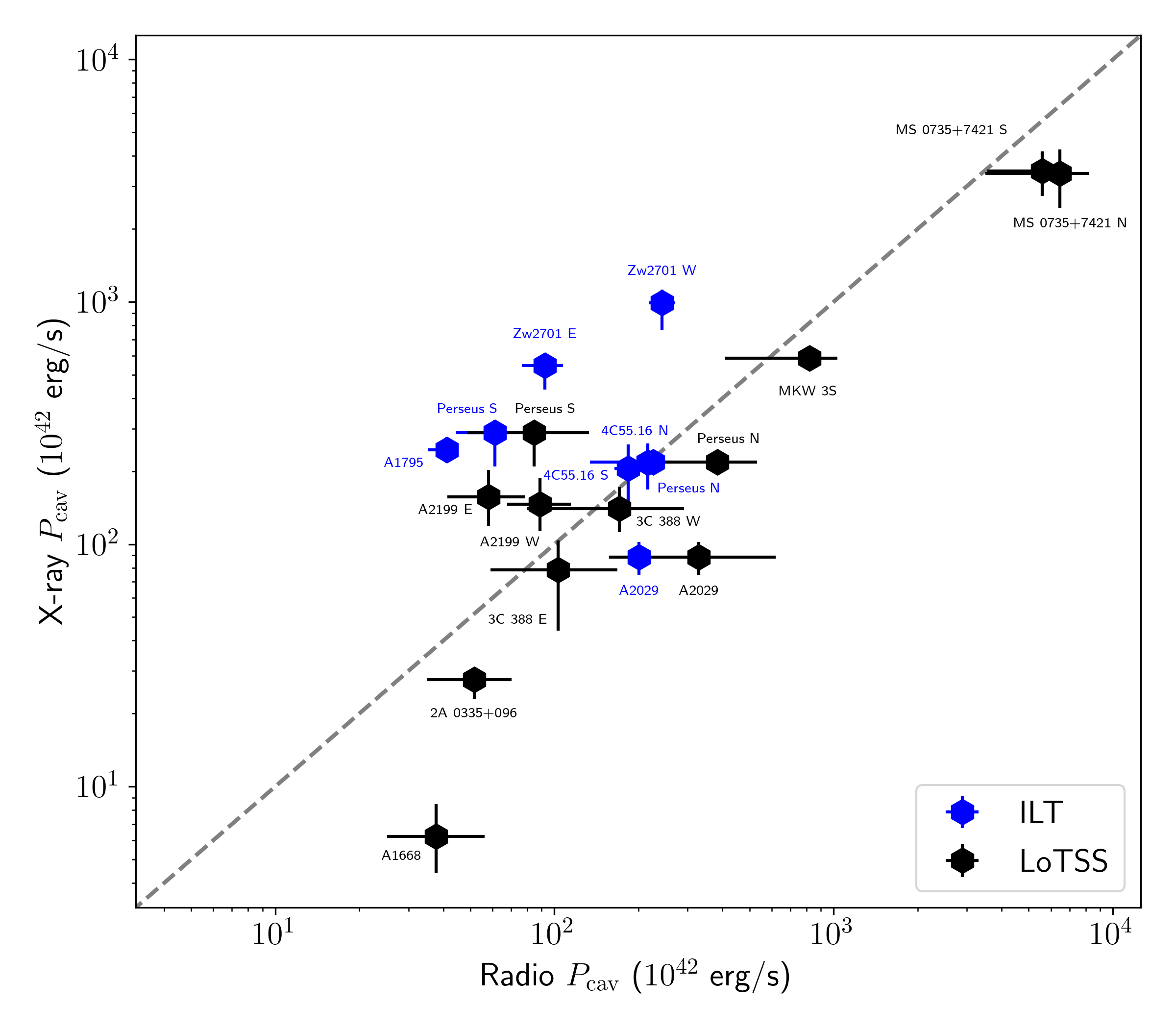}
    \caption{The hybrid X-ray--radio cavity power measurements versus the purely X-ray-based cavity power estimates. The blue data points indicate the measurements derived using high-resolution observations, while the black data points indicate the measurements derived using low-resolution observations.}
    \label{fig:cavitypower}
\end{figure}

For the consistency of the uncertainties on the data, we also calculate the X-ray cavity power and its uncertainty and confirm that these estimates match the published values. The cavity power estimates based only on X-ray observations and the relevant intermediate data are summarized in Table~\ref{tab:xraycavities}. Similarly, the cavity power estimates based on our hybrid X-ray--radio method and the relevant intermediate data are summarized in Table~\ref{tab:radiocavities}. The comparison between the hybrid X-ray--radio measurements and the purely X-ray-based cavity power measurements is shown in Fig.~\ref{fig:cavitypower}.

\begin{figure}
    \centering
    \includegraphics[width=\columnwidth]{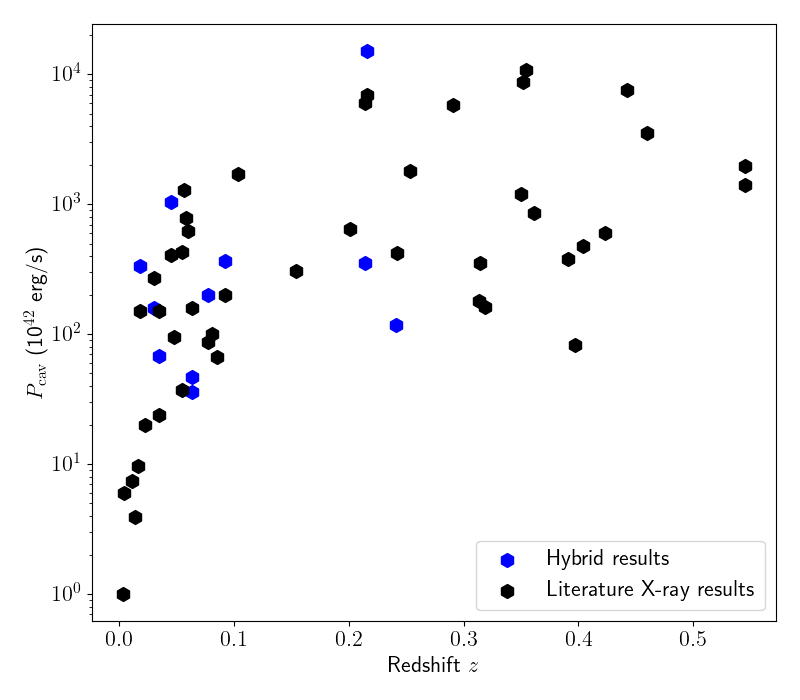}
    \caption{Cavity power as a function of redshift. The blue data points indicate the cavity power measurements derived using our hybrid method, while the black data points indicate the cavity power estimates found in \citet{rafferty06} and \citet{hlavacek12}. From \citet{hlavacek12} we only select the cavities indicated as clearly detected.}
    \label{fig:pcav_z}
\end{figure}

As a final consistency check, we plot our hybrid cavity power estimates as a function of cluster redshift to confirm whether there is any systematic effect between our hybrid estimates and the X-ray-based estimates found in the literature, as shown in Fig.~\ref{fig:pcav_z}. As the two data sets trace the same redshift dependence, it is clear that there is no systematic bias.

\begin{table*}
\caption{Properties of the cavities in our sample derived using the traditional X-ray method.}
    \renewcommand{\arraystretch}{1.2}
    \centering\small
    \begin{tabular}{lllllllll}\hline\hline
        (a)          & (b)   & (c)            & (d)            & (e)   & (f) & (g)             & (h)              & (i)\\
        Cluster name & \(p\) & \(R_\text{l}\) & \(R_\text{w}\) & \(R\) & V & \(t_\text{buoy}\) & \(P_\text{cav}\) & Ref. \\
        & (keV/cm\(^3\)) & (kpc) & (kpc) & (kpc) & (kpc\(^3\)) & (10\(^7\) yr) & (10\(^{42}\) erg/s) & \\\hline
        Perseus (N) & 0.4478 & 8.2 & 4.7 & 6.5 & \(850^{+390}_{-80}\) & \(1.0^{+0.7}_{-0.1}\) & \(218^{+19}_{-26}\) & (1)\\
        Perseus (S) & 0.387 & 9.1 & 7.3 & 9.4 & \(2160^{+610}_{-130}\) & \(1.7^{+1.3}_{-0.3}\) & \(289^{+37}_{-80}\) & (1)\\
        Abell 2199 (E) & 0.1251 & 15 & 10 & 19 & \(7250^{+2890}_{-1650}\) & \(3.1^{+2.2}_{-0.5}\) & \(157^{+46}_{-38}\) & (1)\\
        Abell 2199 (W) & 0.1205 & 16 & 10 & 21 & \(7800^{+3280}_{-1800}\) & \(3.5^{+2.4}_{-0.6}\) & \(146^{+41}_{-33}\) & (1)\\
        2A 0335+096 & 0.1291 & 9.3 & 6.5 & 23 & \(1790^{+670}_{-140}\) & \(5.5^{+3.9}_{-0.9}\) & \(26.6^{+3.2}_{-4.7}\) & (1)\\
        2A 0335+096 & 0.1125 & 4.8 & 2.6 & 28 & \(155^{+71}_{-21}\) & \(10.1^{+6.6}_{-1.4}\) & \(0.98 \pm 0.12\) & (1) \\
        MKW 3S & 0.0675 & 54 & 23 & 59 & \(138200^{+70100}_{-19800}\) & \(9.1^{+5.7}_{-1.2}\) & \(586^{+62}_{-58}\) & (1)\\
        Abell 1795 & 0.242 & 19 & 7.2 & 19 & \(4720^{+2530}_{-580}\) & \(2.8^{+1.7}_{-0.4}\) & \(245^{+32}_{-29}\) & (1)\\
        Abell 1668 (N) & 0.0836 & 2.6 & 1.5 & 3.5 & \(27.8^{+13.7}_{-7.3}\) & \(0.63^{+0.41}_{-0.13}\) & \(2.14^{+0.67}_{-0.53}\) & (2)\\
        Abell 1668 (NW) & 0.0836 & 2.4 & 2.4 & 3.5 & \(57.5^{+20.1}_{-16.1}\) & \(0.67^{+0.50}_{-0.15}\) & \(4.01^{+2.12}_{-1.76}\) & (2)\\
        Abell 2029 & 0.3612 & 13 & 7.2 & 32 & \(3190^{+1450}_{-420}\) & \(7.6^{+5.0}_{-1.1}\) & \(88.3^{+14.1}_{-14.0}\) & (1)\\
        3C 388 (E) & 0.0436 & 15 & 15 & 27 & \(14110^{+2510}_{-2220}\) & \(4.5^{+3.8}_{-0.8}\) & \(78.2^{+25.6}_{-34.2}\) & (1) \\
        3C 388 (W) & 0.0486 & 24 & 10 & 21 & \(12040^{+6130}_{-2860}\) & \(2.3^{+1.5}_{-0.4}\) & \(140^{+33}_{-28}\) & (1) \\
        ZwCl 2701 (E) & 0.7802 & 12.25 & 8.75 & 18.9 & \(4310^{+1520}_{-420}\) & \(3.6^{+2.6}_{-0.5}\) & \(545^{+69}_{-110}\) & (3)\\
        ZwCl 2701 (W) & 0.8364 & 14.0 & 10.5 & 19.25 & \(7030^{+2280}_{-600}\) & \(3.4^{+2.6}_{-0.5}\) & \(993^{+128}_{-228}\) & (3)\\
        MS 0735.6+7421 (N) & 0.0469 & 110 & 87 & 160 & \(3766000^{+954000}_{-472000}\) & \(29.6^{+17.5}_{-5.2}\) & \(3390^{+860}_{-960}\) & (1) \\
        MS 0735.6+7421 (S) & 0.409 & 130 & 89 & 180 & \(4726000^{+1289000}_{-539000}\) & \(32.1^{+17.5}_{-5.2}\) & \(3470^{+700}_{-730}\) & (1) \\
        4C+55.16 (NW) & 0.3135 & 13 & 9.4 & 22 & \(5290^{+1850}_{-620}\) & \(4.4^{+3.2}_{-0.7}\) & \(217^{+44}_{-49}\) & (1)\\
        4C+55.16 (S) & 0.4338 & 10 & 7.5 & 16 & \(2590^{+850}_{-360}\) & \(3.2^{+2.3}_{-0.6}\) & \(205^{+52}_{-55}\) & (1)\\\hline
    \end{tabular}
    \tablefoot{
        Columns: (a) cluster name; (b) ICM pressure; (c) cavity radius along the jet axis; (d) cavity radius perpendicular to the jet axis; (e) cavity distance from the AGN core; (f) cavity volume; (g) buoyancy timescale; (h) cavity power; (i) reference
    }
    \tablebib{(1) \citet{rafferty06}; (2) \citet{pasini21}; (3) \citet{vagshette16}}
    \label{tab:xraycavities}
\end{table*}

\begin{table*}
\caption{Properties of the cavities in our sample derived using the hybrid X-ray--radio method. The horizontal line separates the clusters that are studied at high resolutions (top half) from the clusters that are studied at low resolutions (bottom half).}
    \renewcommand{\arraystretch}{1.2}
    \centering\small
    \begin{tabular}{llllllll}\hline\hline
        (a)          & (b)            & (c)            & (d)   & (e) & (f)             & (g)              & (h)\\
        Cluster name & \(R_\text{l}\) & \(R_\text{w}\) & \(R\) & V & \(t_\text{buoy}\) & \(P_\text{cav}\) & Ref. \\
        & (kpc) & (kpc) & (kpc) & (kpc\(^3\)) & (10\(^7\) yr) & (10\(^{42}\) erg/s) & \\\hline
        Perseus (N) & 6.7 & 5.9 & 6.6 & \(1020^{+190}_{-30}\) & \(1.2^{+1.0}_{-0.2}\) & \(226^{+36}_{-92}\) & (1)\\
        Perseus (S) & 4.3 & 3.4 & 4.5 & \(223^{+64}_{-12}\) & \(0.83^{+0.65}_{-0.13}\) & \(61.3^{+7.6}_{-17.0}\) & (1)\\
        Abell 1795 (N) & 2.2 & 2.3 & 3.5 & \(48.5^{+8.2}_{-6.6}\) & \(0.67^{+0.59}_{-0.11}\) & \(10.5^{+2.7}_{-5.1}\) & (1)\\
        Abell 1795 (S) & 5.5 & 2.3 & 4.7 & \(140^{+72}_{-19}\) & \(0.65^{+0.42}_{-0.09}\) & \(30.7^{+3.1}_{-2.8}\) & (1)\\ 
        Abell 2029 (N) & 9.0 & 3.9 & 10.8 & \(644^{+328}_{-87}\) & \(1.8^{+1.1}_{-0.2}\) & \(76.5^{+7.3}_{-6.8}\) & (1)\\
                       & 9.2 & 5.0 & 31.6 & \(1100^{+510}_{-120}\) & \(8.8^{+5.7}_{-1.2}\) & \(26.1^{+2.4}_{-2.9}\) & (1)\\
        Abell 2029 (S) & 6.1 & 2.7 & 8.2 & \(217^{+107}_{-35}\) & \(1.4^{+0.9}_{-0.2}\) & \(31.9^{+4.1}_{-3.7}\) & (1)\\
                       & 19.5 & 3.9 & 27.1 & \(1480^{+850}_{-210}\) & \(4.7^{+2.9}_{-0.6}\) & \(66.7^{+5.1}_{-4.9}\) & (1)\\
        ZwCl 2701 (E) & 8.4 & 5.6 & 22.3 & \(1220^{+460}_{-170}\) & \(5.8^{+4.1}_{-0.8}\) & \(92.6^{+14.6}_{-16.0}\) & (1)\\
        ZwCl 2701 (W) & 12.7 & 6.3 & 23.3 & \(2390^{+1150}_{-330}\) & \(4.7^{+3.1}_{-0.6}\) & \(243^{+27}_{-25}\) & (1)\\
        4C+55.16 (NW) & 12.6 & 7.4 & 16.0 & \(3270^{+1410}_{-410}\) & \(2.7^{+1.8}_{-0.4}\) & \(216^{+27}_{-29}\) & (1)\\
        4C+55.16 (S) & 13.5 & 6.7 & 20.9 & \(2880^{+1380}_{-400}\) & \(3.9^{+2.5}_{-0.5}\) & \(184^{+21}_{-20}\) & (1)\\\hline
        Perseus (N) & 7.2 & 8.6 & 6.9 & \(2040^{+380}_{-430}\) & \(1.2^{+0.8}_{-0.2}\) & \(384^{+148}_{-189}\) & (1)\\
        Perseus (S) & 4.4 & 4.1 & 4.3 & \(314^{+115}_{-89}\) & \(0.81^{+0.61}_{-0.21}\) & \(84.7^{+48.2}_{-36.1}\) & (1)\\
        Abell 2199 (E) & 12.8 & 6.0 & 16.2 & \(2320^{+1280}_{-730}\) & \(2.6^{+1.7}_{-0.4}\) & \(58.1^{+20.1}_{-16.8}\) & (2)\\
        Abell 2199 (W) & 14.1 & 7.8 & 19.2 & \(4210^{+1990}_{-1090}\) & \(3.2^{+2.1}_{-0.5}\) & \(88.6^{+26.1}_{-21.0}\) & (2)\\
        2A 0335+096 (NW) & 10.2 & 9.4 & 27.0 & \(3960^{+1210}_{-940}\) & \(6.9^{+5.5}_{-1.2}\) & \(40.5^{+17.2}_{-15.6}\) & (3)\\
        2A 0335+096 (SE) & 5.5 & 7.8 & 22.9 & \(1200^{+510}_{-430}\) & \(7.1^{+3.5}_{-1.3}\) & \(11.1^{+7.2}_{-5.8}\) & (3)\\
        MKW 3S & 25.4 & 46.4 & 50.0 & \(201800^{+29700}_{-63300}\) & \(10.0^{+3.7}_{-1.1}\) & \(818^{+213}_{-409}\) & (2)\\
        Abell 1668 (NE) & 13.0 & 7.4 & 31.0 & \(3530^{+2540}_{-1620}\) & \(7.4^{+4.9}_{-1.4}\) & \(21.1^{+14.0}_{-9.6}\) & (2)\\
        Abell 1668 (S) & 9.9 & 9.3 & 34.0 & \(3630^{+2000}_{-1430}\) & \(9.9^{+6.9}_{-2.0}\) & \(16.6^{+12.0}_{-8.1}\) & (2)\\
        Abell 2029 (N) & 21.8 & 4.3 & 21.4 & \(2170^{+3190}_{-1640}\) & \(3.3^{+1.9}_{-0.7}\) & \(132^{+175}_{-100}\) & (2)\\
        Abell 2029 (S) & 26.0 & 4.7 & 22.3 & \(3140^{+4190}_{-2240}\) & \(3.2^{+1.8}_{-0.7}\) & \(197^{+233}_{-140}\) & (2)\\
        3C 388 (E) & 17.3 & 13.7 & 23.3 & \(14900^{+7700}_{-5400}\) & \(3.4^{+2.5}_{-0.8}\) & \(103^{+65}_{-44}\) & (2)\\
        3C 388 (W) & 12.8 & 20.7 & 21.6 & \(19500^{+8300}_{-7300}\) & \(3.3^{+1.5}_{-0.8}\) & \(171^{+120}_{-91}\) & (2)\\
        MS 0735.6+7421 (N) & 111 & 112 & 138 & \(5770000^{+700000}_{-660000}\) & \(23.8^{+16.9}_{-3.8}\) & \(6470^{+1770}_{-2980}\) & (4)\\
        MS 0735.6+7421 (S) & 128 & 115 & 173 & \(7450000^{+920000}_{-780000}\) & \(30.4^{+19.0}_{-4.5}\) & \(5580^{+1310}_{-2030}\) & (4)\\\hline
    \end{tabular}
    \tablefoot{
        Columns: (a) cluster name; (b) cavity radius along the jet axis; (c) cavity radius perpendicular to the jet axis; (d) cavity distance from the AGN core; (e) cavity volume; (f) buoyancy timescale; (g) cavity power; (h) reference for radio observations
    }
    \tablebib{(1) This paper; (2) LoTSS; (3) \citet{ignesti21}; (4) \citet{biava21}}
    \label{tab:radiocavities}
\end{table*}

\section{Discussion}

Constraining the amount of mechanical feedback injected into the ICM by the AGN has not only been considered to be a vital step in understanding the formation and evolution of galaxy clusters, but has also in and of itself been an observational challenge. Although from a physical perspective there is a natural expectation for the radio lobes to paint the same picture as the X-ray cavities, quantifying the amount of mechanical feedback has mainly been performed using X-ray observations. Attempts to enable radio observations to measure the quantity of mechanical feedback have produced significant correlations, but have never made radio observations able to compete with their X-ray counterpart. In this paper, we have described a hybrid method of measuring the quantity of mechanical feedback based on a combination of X-ray and radio observations, and have performed this method on a sample of 14 galaxy clusters for the purpose of verifying whether this new method can be considered to provide reliable results at 144 MHz.

\subsection{General performance}

First of all, we have measured the volume of the radio lobes as observed with LOFAR and compared this volume to the cavity volume estimates derived from X-ray observations. From a physical perspective, there is the expectation that these two volumes should be exactly equal. Therefore, the two measurements must agree within the uncertainties, but that is not always true in our sample. This implies that either the uncertainties are systematically underestimated, or that the simple model that all radio lobes produce clear cavities is invalid. The truth is likely somewhere in the middle. The assumption that the uncertainties on the cavity volume and cavity power are solely determined by projection effects was always known to be oversimplified, but better methods to quantify the uncertainty were lacking. Likewise, the simple "balloon" model in which the radio lobes and ICM are perfectly mutually exclusive is also due a critical review. In reality, additional structures like backflows can make it difficult to distinguish what constitutes as the radio lobe \citep[see, e.g.,][]{timmerman21}. Also, the ICM will mix with the jetted outflows both due to entrainment along the jets as well as due to turbulence within the radio lobes. This underlines the importance of high-quality observations. For future practical purposes, it will be most useful to assume the scatter we observe between our radio-derived cavity measurements and the X-ray cavity measurements (\(\sigma=0.30~\mathrm{dex}\)) as a systematic uncertainty on these measurements. For direct comparison with literature results, we maintain the projection-derived uncertainties for the remainder of this paper.

Using the volume of the cavities derived based on LOFAR observations in combination with the X-ray-derived ICM pressure, we derive the cavity power of the AGN. In general, this method produces a relatively tight correlation with the purely X-ray-derived estimates, with a scatter in the cavity power measurements of only 0.44~dex. This is highly competitive with previous attempts to derive a correlation between the radio properties and the cavity power in galaxy clusters, which resulted in scatters of at least 0.8~dex based on monochromatic radio powers \citep{birzan08, kokotanekov17} and a scatter of 0.65~dex based on bolometric radio luminosities of only the radio lobes \citep{birzan08}. In addition, there appears to be no significant systematic difference between the hybrid X-ray--radio method and the purely X-ray-based method. However, it is valuable to explore deeper into the results and investigate exactly how the hybrid X-ray--radio method performs under different circumstances.

In general, the best correlation is seen in systems where both the X-ray cavities are clearly detected and their sizes can be well constrained. If we classify our sample using the figure of merit (FOM) system from \citet{rafferty06}, where cavities fully surrounded by bright rims score a 1, cavities partially surrounded by bright rims score a 2, and cavities with either a faint rim or no rim score a 3, we can quantify how strong this effect is. We assume the FOMs published by \citet{rafferty06} for this, and classify Abell\,1668 (FOM=3) and ZwCl\,2701 (FOM=2) ourselves using the same criteria, as for these systems we use the cavity measurements from \citet{pasini21} and \citet{vagshette16}, respectively. There is only one system classified to have a FOM of 1 (Perseus), which we will discuss in more detail later. Comparing the cavity power measurements of cavities with FOM=2 (N=10) and FOM=3 (N=6), we find that the FOM=2 systems show a scatter of 0.34 dex, while the FOM=3 systems show a scatter of 0.49 dex instead. This suggests that in some instances, using the radio lobes to measure the cavity volume can provide a better estimate of the cavity power than the X-ray cavities.

By comparing the cavity power estimates as a function of redshift between the traditional method and the hybrid method, we confirm that the resulting distributions are in good agreement, allowing the hybrid method to be used to study a sample of galaxy clusters. This is particularly important as this enables the hybrid method to be confidently used at higher redshifts as well, where the X-ray cavities can in general not be detected due to sensitivity limitations. In the event that the hybrid method is applied to clusters at higher redshifts, where the X-ray observations may not be able to aid with the identification of the radio lobes, the uncertainties on the dimensions of the radio lobes may increase, depending on the exact morphology and brightness. However, as the radio lobes are in general clearly identifiable, this is not expected to cause significantly increased uncertainties in general. This also applies to clusters where the central ICM pressure is obtained based on Sunyaev-Zel'dovich measurements instead of X-ray observations.

At higher redshifts, the surface brightness of the radio lobes will naturally decrease, resulting in a soft limit on measurements of the cavity power, also depending on the scale of the lobes. However, clear detection of the lobes in the radio galaxy 4C\,43.15 at \(z=2.4\) by \citet{sweijen22} and further detections of the radio lobes in the protoclusters 4C\,41.17 (\(z=3.8\)), B2\,0902+34 (\(z=3.4\)) and 4C\,34.34 (\(z=2.4\)) by Cordun et al. (in prep.) provide an encouraging perspective and demonstrate the feasibility of detecting radio lobes with the ILT even at high redshifts.

\subsection{Performance per system}\label{sec:performance}

To better understand how the method performs, it is useful to consider the separate systems in more detail. For three systems (RX\,J1532.9+3021, MACS\,J1720.2+3536 and MACS\,J1423.8+2404), the radio observation did not reveal sufficient detail and structure to be able to derive a radio lobe volume. This suggests that the hybrid X-ray--radio method is mostly viable for sources brighter than \({\sim}100\)~mJy at 144~MHz. Such a limit on radio brightness is likely to introduce selection effects, especially towards higher redshifts, so it is important to be aware of this.

Proceeding with the sources which are well detected in our LOFAR observations, one of the most interesting examples is the Perseus cluster. Whereas the cavity power predicted by the X-ray method and the hybrid X-ray--radio method agree very well for the Northern lobe, the Southern lobe shows a significant difference. In the radio map, the southern lobe appears to consist of two distinct components: a bright compact component directly South of the AGN and a faint extended component towards the South-West of the AGN. Comparing our LOFAR map with deep VLA imaging at 1.5~GHz \citep{gendronmarsolais21} reveals a notable difference in the spectral index, with the bright compact component featuring an average spectral index of \(\alpha=-1.2\) and the faint extended component featuring a spectral index of \(\alpha=-1.7\). This suggests that the extended component is much older and likely corresponds to a previous outburst of the AGN. As the Perseus cluster is the lowest-redshift cluster in our sample, the faint component can be reasonably well detected in our radio maps, but this would likely not hold at higher redshifts. In the X-ray map, the two components are difficult to distinguish, leading to the cavity volume being estimated based on the combination of the two components. If the second component can not be clearly detected using radio observations, this would lead to a significant discrepancy between the resulting cavity power measurements. We note that this situation can occur in any system where an old episode of AGN activity can still be traced through its cavity in the ICM, as the magnetized plasma of the radio lobe will only remain visible at radio wavelengths for a limited period. Synchrotron-emitting cosmic-ray electrons generally experience lifetimes of \(\lesssim 10^8\)~years \citep{feretti12, vanweeren19}, which may cause these to decay within the lifespan of older ICM cavities (see Tables~\ref{tab:xraycavities} and \ref{tab:radiocavities}).

Based on the radio morphology of the jetted outflows of the AGN, we note that our assumption that the radio lobes propagate radially outwards from the center of the cluster is not always strictly true. In the case of Abell 2029, it is particularly clear that the jetted outflows can bend away from their initial direction, but similar structures are also present in Abell 1795 and MS 0735.6+7421. Because of this jet bending, the assumption that the age of a cavity is only a function of the radial distance starts to fail. However, we estimate that this effect is relatively negligible in comparison to the other contributions to the overall uncertainty on the cavity power measurements.

In two instances (Abell\,1795 and 2A\,0335+096) we observe that the ICM as observed in the X-ray regime features such complex structures that this affects the identification and description of the cavities. Unless the cavities are very clearly present in the image, they are generally identified by subtracting a smooth model of the ICM brightness profile from the image, which causes the cavities to appear as negative regions. However, the presence of additional structure can affect the fitting of a brightness profile and thereby result in unreliable cavity detections. In these instances, relying on the volume of the radio lobes may be preferred over the X-ray cavity volume.

Similarly, there are clusters where even after model subtraction, the cavities are not apparently obvious among the residuals. For example, both the Abell\,1668 and Abell\,2029 systems have been reported to feature cavities, but only at low significance (\citealt{rafferty06} and \citealt{pasini21}, respectively). This demonstrates that even in more relaxed clusters, the radio lobes may provide the most accurate estimates of the cavity sizes.

\subsection{Comparing low- and high-resolution imaging}

As two sources in our high-resolution sample (Perseus and Abell\,2029) are sufficiently extended to also be studied at an angular resolution of \({\sim}6\)~arcseconds, we are able to compare the effect of angular resolution on the cavity power estimates. In principle, the expectation is that there should be no dependency on angular resolution assuming that the deconvolved size of the radio lobe is estimated. However, we do observe that the low-resolution maps in general result in a higher cavity power estimate.

In the case of Abell\,2029, our high-resolution map reveals a complex structure which is difficult to recover from the low-resolution map. We expect that the lobe volume is overestimated in our low-resolution map as the complex structure is smoothed out. The Perseus cluster is a different case, as both the Northern and the Southern lobes appear smooth and ellipsoidal in shape. For the sake of consistency and comparison, we have chosen to only consider the bright compact component south of the AGN instead of including the faint extended component towards the southwest. This sustains the discrepancy observed with the X-ray cavities but allows for a more direct comparison between the high-resolution and low-resolution maps. It is uncertain why the low-resolution map indicates a larger lobe volume, but it is likely that a contribution from the diffuse mini halo \citep{soboleva83} blends with the outer edge of the radio lobes. In general, we give preference to high-resolution observations unless the reduced surface brightness sensitivity of such observations results in a poorly constrained radio lobe volume.

\section{Conclusions}

We have described and tested a hybrid method for measuring the cavity power as an estimate of the amount of mechanical feedback injected by AGNs into their environment. This method is based on a combination of X-ray and radio observations, where the X-ray supplies most of the environmental parameters and the radio observations are used to determine the volume of the cavities in the ICM. By testing this method on a sample of 14~clusters and comparing the hybrid method to the traditional X-ray-based method, we find that the radio-derived cavity volumes are in good agreement with the X-ray-derived cavity volumes, although the systematic uncertainties are in general likely to be underestimated. After calculating the cavity powers associated with the observed cavity volumes, we measure a scatter of 0.44~dex on the correlation between the traditional and the hybrid method, and this scatter improves as the cavities are more clearly detected in the X-ray observations. Thanks to the LOFAR long baselines, the combination of sensitivity for diffuse radio lobes and the angular resolution to tightly constrain the cavity volume is available for the first time, enabling radio-mode feedback to be studied reliably even at high redshifts. As demonstrated by the number of unique ILT observations processed for this paper, the hybrid method can feasibly be used on relatively large samples of clusters.

From further analysis, we note that the radio observations can in general only be used on sufficiently bright sources of at least \({\sim}\)100~mJy, and recommend careful consideration for the presence of old steep-spectrum plasma which may fall below the sensitivity limit even at low frequencies. Likewise, we also see instances where the radio lobes may be considered to provide the most accurate estimate of the cavity volume as the X-ray cavities are not reliably detected. In general, we recommend that the choice of method is made per cluster based on the quality and contents of the available data.

\begin{acknowledgements}
      We would like to thank the anonymous referee for useful comments. R.T. and R.J.v.W. acknowledge support from the ERC Starting Grant ClusterWeb 804208. A.B. acknowledges support from the VIDI research programme with project number 639.042.729, which is financed by the Netherlands Organisation for Scientific Research (NWO). This paper is based (in part) on data obtained with the International LOFAR Telescope (ILT) under project code LC14\_019. LOFAR \citep{haarlem13} is the Low Frequency Array designed and constructed by ASTRON. It has observing, data processing, and data storage facilities in several countries, that are owned by various parties (each with their own funding sources), and that are collectively operated by the ILT foundation under a joint scientific policy. The ILT resources have benefitted from the following recent major funding sources: CNRS-INSU, Observatoire de Paris and Université d'Orléans, France; BMBF, MIWF-NRW, MPG, Germany; Science Foundation Ireland (SFI), Department of Business, Enterprise and Innovation (DBEI), Ireland; NWO, The Netherlands; The Science and Technology Facilities Council, UK; Ministry of Science and Higher Education, Poland. The National Radio Astronomy Observatory is a facility of the National Science Foundation operated under cooperative agreement by Associated Universities, Inc. The Jülich LOFAR Long Term Archive and the German LOFAR network are both coordinated and operated by the Jülich Supercomputing Centre (JSC), and computing resources on the supercomputer JUWELS at JSC were provided by the Gauss Centre for Supercomputing e.V. (grant CHTB00) through the John von Neumann Institute for Computing (NIC).
\end{acknowledgements}

\bibliographystyle{aa}
\bibliography{refs}

\begin{appendix}

\section{Spectal index maps}\label{appendix:vla}

\begin{figure*}
    \centering
    \includegraphics[width=\textwidth]{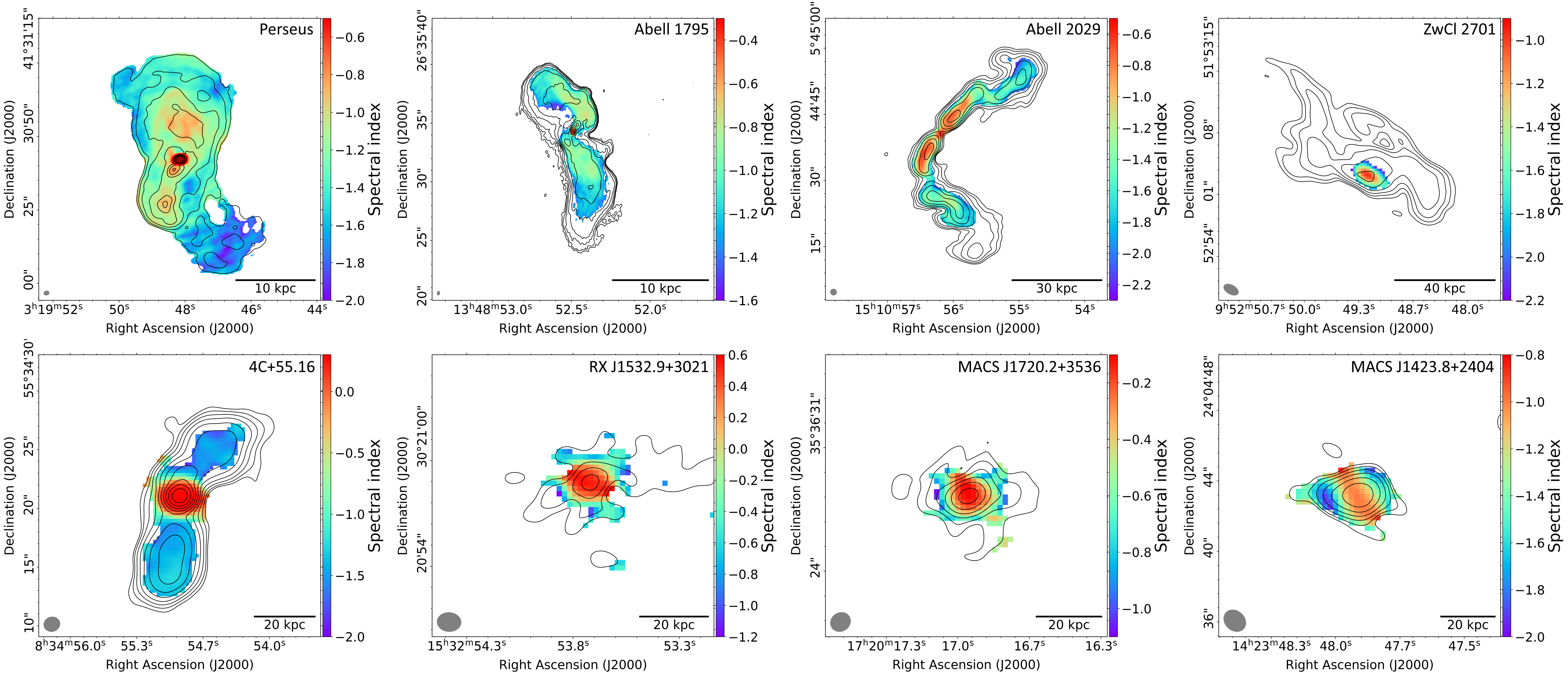}
    \caption{Spectral index maps of the high-resolution sample based on LOFAR and VLA observations. The spectral index for all targets is calculated between 144 MHz and 1.5 GHz, except for Abell 1795, for which the spectral index is calculated between 144 MHz and 8 GHz. All spectral index maps are masked below 3\(\sigma\) confidence. The black contours indicate the radio intensity at 144 MHz and are drawn in increments of 2, starting at 5 times the rms noise level. The beam is indicated in grey in the bottom left corner of each panel.}
    \label{fig:spixmaps}
\end{figure*}

To produce spectral index maps of our high-resolution sample, we obtained archival VLA observations at 1.5 GHz (Project codes: AA54, AK145, AJ99, AS309, AC243, AL252, AR343, 14A-040) for most of the sources in our high-resolution sample. The 1.5 GHz VLA map of Perseus was obtained courtesy of \citet{gendronmarsolais21}, and for Abell 1795 we opted to obtain the 8 GHz observations (Project codes: AG262, AG273). All raw data were processed following the standard VLA data reduction procedure \citep[e.g.,][]{timmerman20}. Next, for each cluster, we applied the same \(uv\) lower limits to both the LOFAR and VLA data and smoothed the resulting images to the same synthesized beam. Finally, we obtain the spectral index maps shown in Figure \ref{fig:spixmaps}.

\end{appendix}

\end{document}